\documentclass[pre, twocolumn, showpacs, superscriptaddress, showkeys, longbibliography]{revtex4-1}
\usepackage{graphicx}
\usepackage{bm}
\usepackage{amssymb, amsmath, amsthm, amsfonts}
\usepackage{hyperref}
\usepackage{xcolor,soul}
\begin{document}


\title{Steady State of an Active Brownian Particle in Two-Dimensional Harmonic Trap}
\date{\today}

\author{Kanaya Malakar}
\affiliation{Presidency University, 86/1, College Street, Kolkata 700073, India}
\affiliation{Martin A. Fisher School of Physics, Brandeis University, Waltham, Massachusetts 02453, USA}
\author{Arghya Das}
\author{Anupam Kundu}
\author{K. Vijay Kumar}
\author{Abhishek Dhar}
\affiliation{International Centre for Theoretical Sciences, Tata Institute of Fundamental Research, Bengaluru 560089, India}

\begin{abstract}
We find an exact series solution for the steady-state probability distribution of a harmonically trapped active Brownian particle in two dimensions, in the presence of translational diffusion. This series solution allows us to efficiently explore the behavior of the  system in different parameter regimes. Identifying ``active'' and ``passive'' regimes, we predict a surprising re-entrant active-to-passive transition with increasing trap stiffness. Our numerical simulations validate this finding. We discuss various interesting limiting cases wherein closed form expressions for the distributions can be obtained. 
\end{abstract}



\maketitle

\section{Introduction}
The study of active particles has seen an upsurge of interest in recent years for its relevance in describing 
many non-equilibrium processes \cite{Romanczuk_2012_review,bechinger_active_2016}. Active Brownian particles (ABPs) and run-and-tumble particles (RTPs) are minimal models for self-propelled active particles \cite{Cates2013}. The dynamics of such active particles breaks detailed balance at the microscopic level, and is characterized by properties which are remarkably different from equilibrium systems. Systems composed of interacting active particles are known to exhibit a plethora of exotic phenomenon including flocking~\cite{Toner_2005_flocking,Kumar_2014_flocking}, clustering~\cite{Slowman_2016_clustering,Slowman_2017_clustering}, motility induced phase separation and segregation~\cite{Cates_2015_motility,Patch_2017_motility,Redner_2013_motility,Stenhammar_2013_motility}, ratchet effects~\cite{Reichhardt_2017_ratchet} etc.

While there exists a considerable body of work on the hydrodynamic description of active matter \cite{Marchetti2013}, the number of exactly solvable models that illuminate the novel statistical physics of active particles are few. A number of recent studies, both experimental \cite{li2009accumulation,maggi2014generalized,Takatori_2016_exptAcousticTweezer,Walsh_2017_exptSingleParticle,Dauchot_2018_exptHexbug} and theoretical \cite{elgeti2015run,Prymidis,Jahanshahi2017,wagner2017steady,pototsky,Duzgun_2018_2D,Basu_2018_2D,das2018confined}, show that even a single active particle can exhibit rich and counter-intuitive physics such as non-Boltzmann distributions peaked away from potential minima \cite{Takatori_2016_exptAcousticTweezer,Dauchot_2018_exptHexbug,dhar2018} and clustering \cite{tailleur2009sedimentation}. For example, a recent experimental study of active Janus particles, in a two dimensional effectively harmonic trap, surprisingly observed that, in the dilute limit, the trap stiffness can be tuned to induce a crossover in the particle distribution from a Boltzmann-like distribution peaked at the trap center to a strongly active non-Boltzmann distribution with off-centered peaks \cite{Takatori_2016_exptAcousticTweezer}. At first sight, the dynamics of a single active particle appears to be a small variation of passive Brownian motion. However, one finds that calculating even the steady state probability distribution is highly non-trivial. Some exact results obtained in  \cite{hanggi11995,Solon_2015_pressure,Malakar_2018_1D,Demaerel_2018_1D,dhar2018,mallmin2018exact,elgeti2015run,wagner2017steady,pototsky,Duzgun_2018_2D,Basu_2018_2D,das2018confined} indicate the qualitatively rich physics that even single particle active systems can exhibit. Exactly solvable models of isolated active particles are thus important not only for understanding laboratory experiments of self-propelled particles in confined geometries, but they also provide a good starting point to study systems of weakly interacting active particles.

In this paper, we consider the stochastic dynamics of an active Brownian particle (ABP) confined to move in an isotropic potential. Our main result is an exact series solution of the corresponding Fokker-Planck equation in the steady-state. From this exact solution, we find that the radial probability distribution of the ABP has two ``phases'': a ``passive phase'' described by a Boltzmann-like distribution with a peak at the trap center, and an ``active phase'' described by a non-Boltzmann distribution where the probability distribution is peaked away from the trap center. We find that increasing the trap stiffness induces a transition from the passive phase to the active phase as experimentally observed in \cite{Takatori_2016_exptAcousticTweezer}. Our exact solution reveals a surprising prediction of a re-entrant transition to the passive phase upon stiffening the harmonic trap further. Note that these ``transitions'' are not thermodynamics phase-transitions, but are crossover behaviors in the probability distribution for the ABP. We estimate typical parameter values where such a re-entrant transition can be experimentally observed. Furthermore, the exact solution in our study unifies the various asymptotic limits of the steady-state distribution considered in earlier studies. Our numerical analysis is in good agreement with the analytical results.

The plan of the paper is as follows:- In Sec.~\eqref{sec:results} we define the precise model and then present a  summary of our main analytical results for the steady state distribution, along with their numerical verification. The details of the analytical computations are presented in the appendices. In Sec.~\eqref{sec:limits} we discuss some special limiting cases and finally we conclude with a discussion of our results in Sec.~\eqref{sec:summ}. 

\section{Trapped noninteracting ABPs: Steady state solution}
\label{sec:results}
The two-dimensional motion of  an ABP in an isotropic harmonic potential ($k\rho^2/2$), with position coordinate ${\bm \rho}=(\rho \cos \varphi, \rho \sin \varphi)$  and internal angular degree of freedom $\theta$ at time $\tau$, are governed by the Langevin equations:
\begin{align}\begin{split}
\frac{d { \bm \rho}}{d \tau} &= u_0 \hat{{\bf e}} (\theta) - \mu k {\bm \rho} + \sqrt{2D_{t}} ~ {\bm \xi}_r(\tau), \\
\frac{d \theta}{d \tau} &= \sqrt{2 D_{\theta}} ~ \xi_{\theta}(\tau),\end{split} \label{eq2}
\end{align}
where $\mu$ is the translational mobility. The ABP self-propels along the direction $\hat{\bf e} (\theta)=(\cos \theta, \sin \theta)$ with speed $u_0$. The Gaussian random variables ${\bm \xi}_r(\tau)$ and $\xi_\theta(\tau)$, with zero mean and unit variance, 
are uncorrelated in time, and represent translational and rotational noise terms. 
Defining  dimensionless variables $ {\bf r} = {\bm \rho} \sqrt{{D_{\theta}}/{D_t}}$ and $t = \tau D_{\theta}$,  the Langevin equations take the  form 
\begin{align}\begin{split}
\frac{d{\bf r}}{dt} &= \lambda \hat{{\bf e}} (\theta) - \beta {\bf r} + \sqrt{2} ~ {\bm \xi}_r(t),  \\ 
\frac{d \theta}{dt} &= \sqrt{2} ~ \xi_{\theta}(t), \end{split} \label{eq3}
\end{align} 
where
$\lambda = {u_0}/{\sqrt{D_{\theta} D_t}}$ and $\beta = {\mu k}/{D_{\theta}}$. Note that $\lambda$ is a P\'eclet number that indicates the relative importance of persistent motion compared to diffusion. Our aim is to calculate the steady state distribution of the trapped ABP. The Fokker-Planck equation for the probability distribution $\mathcal{P}(r,\varphi,\theta,t)$, corresponding to the above Langevin equation, can be obtained using standard methods \cite{Risken}.  For the case of a  free ABP in two dimension this was obtained in \cite{Sevilla2015}. In our case the corresponding Fokker-Planck equation  is
\begin{eqnarray}
\frac{\partial \mathcal{P}}{\partial t} = -{\bf \nabla} \cdot [(\lambda \, \hat{{\bf e}}(\theta) -\beta \, \mathbf{r} -{\bf \nabla})\mathcal{P}] 
+ \frac{\partial^2 \mathcal{P}}{\partial \theta^2}~. \label{eq6}
\end{eqnarray}
Note that the the radial symmetry of the problem implies that the steady state joint probability distribution $\mathcal{P}(r,\varphi,\theta) \equiv \mathcal{P}(r,\chi)$ depends only on the combination $\chi = \theta - \varphi$. In the steady state ($\partial_t \mathcal{P}=0$), the Fokker-Planck equation can be recast in the following form
\begin{align}
\mathcal{L}_0 \mathcal{P} &= \lambda \, \mathcal{L}_1 \mathcal{P}, \label{eq14}
\end{align}
where $\mathcal{L}_0$ is an operator which is second-order in $r$ and $\chi$ and contains the terms arising from the confining potential. On the other hand $\mathcal{L}_1$  is a first-order operator that arises from the drift-terms in \eqref{eq6}.  Their explicit forms are given by
\begin{align}
\mathcal{L}_0 &=  \frac{1}{r} \frac{\partial}{\partial r} \bigg[ r \bigg( \frac{\partial}{\partial r} + \beta r \bigg) \bigg]
+ \frac{1}{r^2} \frac{\partial^2 }{\partial \chi^2} + \frac{\partial^2 }{\partial \chi^2}~, \label{L0} \\
 \mathcal{L}_1 &= \cos \chi \frac{\partial}{\partial r} - \frac{\sin \chi}{r} \frac{\partial}{\partial \chi}~. \label{L1}
\end{align}

 We propose to solve Eq.~\eqref{eq14} in the form of a power-series expansion in $\lambda$: 
\begin{align}
\mathcal{P}(r, \chi) = \sum_{m=0}^\infty \lambda^{m} \mathcal{P}^{(m)}(r,\chi). \label{P-joint}
\end{align}
Using this ansatz in \eqref{eq14} and equating powers of $\lambda$ on both sides, we get 
$\mathcal{L}_0 \mathcal{P}^{(0)} = 0$ and $\mathcal{L}_0 \mathcal{P}^{(m)} = \mathcal{L}_1 \mathcal{P}^{(m-1)}$ for $m \geq 1$.
The zeroth-order solution is $\mathcal{P}^{(0)} (r, \chi) = e^{-\beta r^2/2}/{Z}$, where $Z=2 \pi \int_0^\infty dr \, r e^{-\beta r^2/2}$.  Note that this is identical to the equilibrium distribution corresponding to the potential $V(r)=r^2/2$, if $D_t=\mu k_BT$. The solution at the next order satisfies $\mathcal{L}_0 \mathcal{P}^{(1)} = \mathcal{L}_1 \mathcal{P}^{(0)}$.  Thus, if the eigensystem of $\mathcal{L}_0$ is explicitly known, then the source term $\mathcal{L}_1 P^{(0)}$ (obtained from the previous order) can be expanded in the eigenbasis of $\mathcal{L}_0$. This allows one to solve for $\mathcal{P}^{(1)}$. Following this strategy at every order $m$, we find (see  App.~\ref{Appendix-A}) that the solutions $\mathcal{P}^{(m)}$ can be written as
\begin{equation}
\mathcal{P}^{(m)}(r, \chi) = \sum_{n,l} C_{n, l}^{(m)} \phi_{n,l}(r, \chi). \label{p^m}
\end{equation}
where the summation on $n$ and $l$ is constrained by $m=2n + |l|$ and
\begin{eqnarray}
\phi_{n,l}(r, \chi) &=& \bigg[ \frac{n! (\frac{\beta}{2})^{|l|+1}}{\pi \Gamma(n+|l|+1)} 
\bigg]^{\frac{1}{2}} r^{|l|} e^{-\frac{\beta r^2}{2}} L_n^{|l|} \bigg(\frac{\beta r^2}{2}\bigg)  e^{il \chi}~, \nonumber
\end{eqnarray}
is the right eigenfunction of $\mathcal{L}_0$ corresponding to the eigenvalue $\nu_{n,l} = -\beta  (2n + |l|)-l^2$ with $n \geq 0$, $l$ being integers, $\Gamma$ is the gamma function, and $L_n^{|l|}(x)$ is the generalized Laguerre polynomial. The expansion coefficients $C_{n, l}^{(m)}$ satisfy the following recursion relations
\begin{align}
C^{(m)}_{n,l} &=  \frac{C^{(m-1)}_{n,l-1} \sqrt{(n+|l|) \frac{\beta}{2}} - C^{(m-1)}_{n-1,l+1} 
\sqrt{n \frac{\beta }{2}} }{\beta  (2n + |l|) + l^2}~,~l>0 \nonumber \\
C^{(m)}_{n,0} &= -{C^{(m-1)}_{n-1,1}}/{\sqrt{2 \beta n}}~.  
\label{recursion-main}
\end{align}
with $C^{(0)}_{0,0}=1/\sqrt{2 \beta}$. Solving the above recursion relations, one can, in principle, evaluate $\mathcal{P}^{(m)}$ at any order $m$. For instance, the explicit form of $\mathcal{P}^{(m)}$ for the first few orders are,
\begin{align}
\begin{split}
\mathcal{P}^{(0)}  &= \frac{\beta }{2 \pi} e^{- \frac{\beta r^2}{2}},~~\mathcal{P}^{(1)} = \frac{\beta^2 r \cos\chi}{2 \pi (\beta+1)}  e^{- \frac{\beta r^2}{2}} \\
\mathcal{P}^{(2)}  &= \frac{ \beta ^3 r^2 \cos (2 \chi )+ \beta(\beta +2)
   (\beta r^2-2)}{8 \pi  (\beta+1) (\beta +2)} e^{-\frac{\beta r^2}{2}. }
\end{split}
\label{P-0-2}
\end{align}
Expressions for few more orders are given in App.~\ref{Appendix-B}.
The analytical evaluation of $C^{(m)}_{n,l}$ becomes tedious with increasing order of $m$. However, they are easily computed numerically. Thus one can evaluate the joint distribution $\mathcal{P}(r,\chi)$ up to any power of~$\lambda$.  Integrating \eqref{P-joint} over the angular variable $\chi$, only $l=0$ terms contribute and we find that the marginal radial distribution function is given by,
\begin{eqnarray}
P(r) = \sum_{n=0}^{\infty} \lambda^{2n} \sqrt{2 \pi \beta}~ C_{n,0}^{(2n)} e^{-\frac{\beta r^2}{2}} 
L_{n} (\beta r^2/2),~ \label{radPDF}
\end{eqnarray}
where $L_n(z)$ is $n^{\rm th}$ Laguerre polynomial and the coefficients $C_{n,0}^{(2n)}$ can be computed numerically easily from the recursion relations in \eqref{recursion-main}. From numerics, we found that the ratio of two consecutive $C^{(2n)}_{n,0}$ decreases as $\sim n^{-1}$ for large $n$, which ensures that the series in Eq.~(\ref{radPDF}) is convergent.

\begin{figure}[t]
\includegraphics[width=\linewidth]{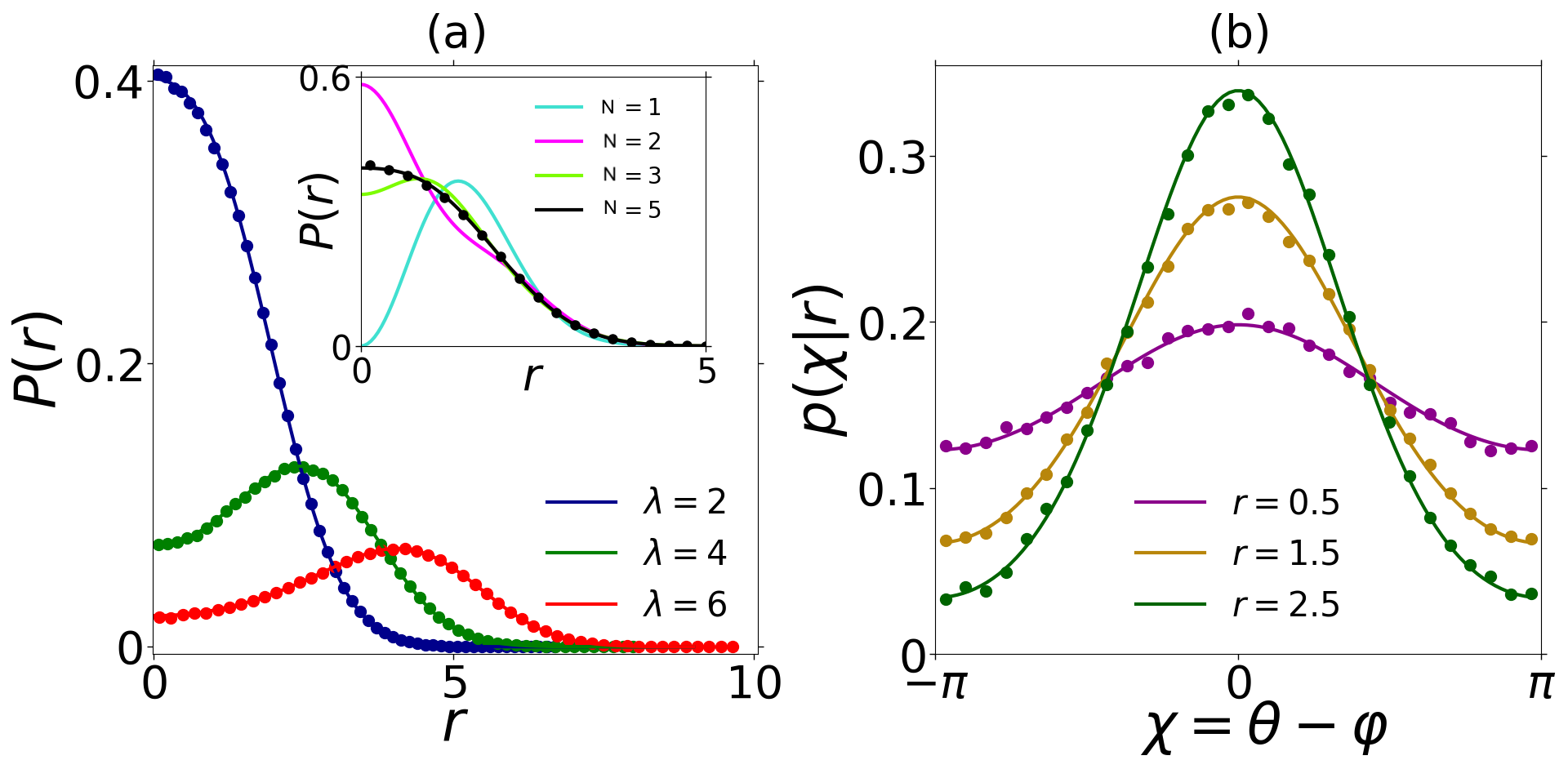}
\caption{
(a) Comparison of the exact solution (solid lines) Eq. \eqref{radPDF} with direct numerical simulations (disks) of the Langevin equations for the ABP. Notice the shift in the peak of the distribution with increasing $\lambda=u_0/\sqrt{D_t D_{\theta}}$. The inset shows the convergence of the exact solution (for $\lambda=2$) as more terms ($M$) are included in Eq. \eqref{radPDF} to evaluate $P(r)$. (b) The conditional distribution $p(\chi|r)$ at various $r$ compared with simulations for $\lambda=1$. The peaks at $\chi=0$ indicate a radial orientation of the ABPs far away from the trap center.}
\label{P-for-gen-lambda}
\end{figure}

We numerically solved Eqs.~(\ref{eq2}) using the Euler-Maruyama method \cite{sde} using a time-step $\Delta t=10^{-4}$ and averaged over $10^6$ realizations. Our analytical solution of the Fokker-Planck equation is in good agreement with direct numerical simulation results of  Eqs.~(\ref{eq2}). This is shown in Fig.~(\ref{P-for-gen-lambda}a)  where we compare the analytical result for $P(r)$ in Eq.~\eqref{radPDF} (summed over a finite number of terms, $N$) with  those from the simulations. The inset of Fig.~(\ref{P-for-gen-lambda}a) shows the convergence of the series with increasing $M$. We find excellent agreement even for $M=5$. 

\begin{figure}[t]
\includegraphics[width=\linewidth]{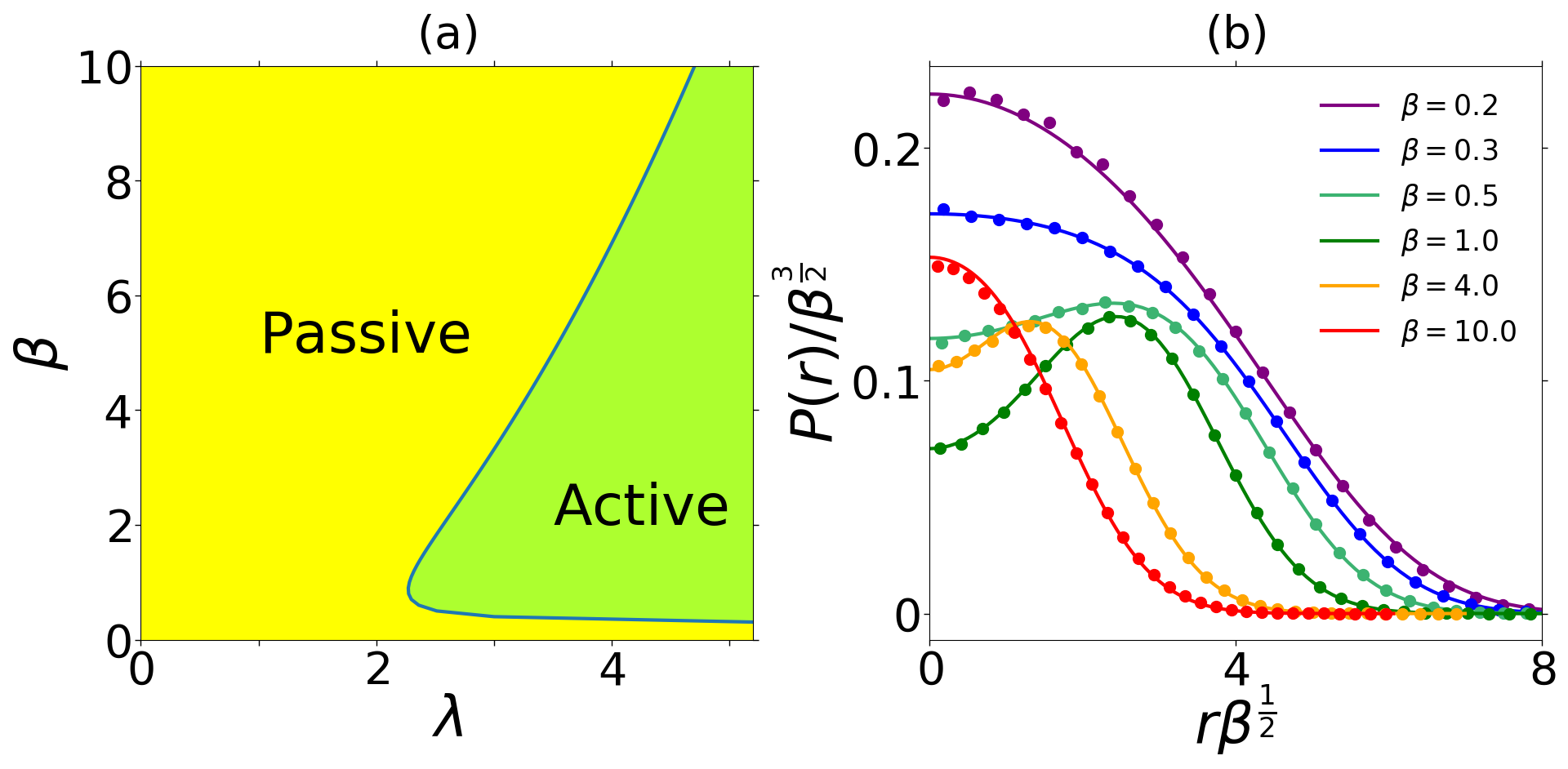}
\includegraphics[width=\linewidth]{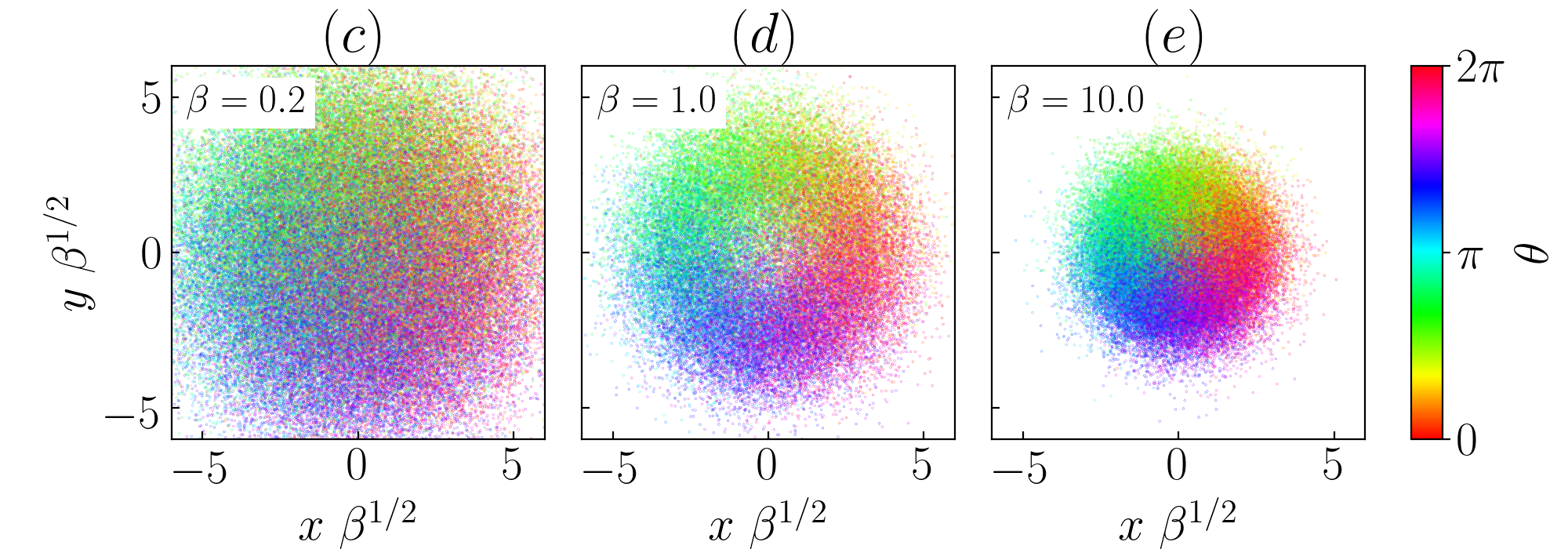}
\caption{
(a) Active and passive regions of the ABP distribution Eq.\eqref{radPDF} in the $\lambda-\beta$ plane. The solid line demarcating the active and passive regions is given by $P''(r) \vert_{r=0}=0$. In the passive region, the distribution peak coincides with the trap center ($r=0$), while in the active region the peak of $P(r)$ is away from $r=0$. Notice that for a high-enough activity (large $\lambda$), increasing the trap-stiffness (proportional to $\beta$) first drives a passive-to-active transition and then a reentrant active-to-passive transition. (b) This reentrant transition predicted by the analytical result Eq. \eqref{radPDF} (solid lines), summed upto a $M=150$ terms, is confirmed by our numerical simulations (disks) at increasing values of $\beta$ at fixed $\lambda=4$. In (c-e), we show the steady-state probability distribution $P(\mathbf{r}, \theta)$ as a cloud of points each colored by its values of $\theta$ at a fixed value of $\lambda=4$. In the passive phase, $\beta=0.2$, we observe that the probability is peaked at the center of the trap while $\theta$ is randomly distributed. In the active phase, $\beta=1$, $P(\mathbf{r}, \theta)$ is peaked away from the trap-center while the distribution of $\theta$ is correlated with the polar angle $\phi$. Finally, in the re-entrant passive phase, $\beta=10$, although the distribution of points is peaked at the trap-center, we observe a clear pattern in the distribution of $\theta$ similar to the active phase in $(d)$.}
\label{crossover}
\end{figure}

The distribution $p(\chi)$, of the relative orientation $\chi$, can be obtained by integrating $r$ from the joint distribution $\mathcal{P}(r,\chi)$ (given in Eq.~\eqref{P-joint}). However, it is more interesting to look at the conditional distribution $p(\chi |r)=\mathcal{P}(r,\chi)/P(r)$ of the orientation $\chi$ at a given $r$. We observe from Fig.~(\ref{P-for-gen-lambda}b) that $p(\chi |r)$ is peaked at $\chi=0$. With increasing $r$, the width of this peak decreases while its height increases.  This indicates that the particles which successfully climb the potential must have their orientation $\hat{\bm e}(\theta)$ preferentially directed along the radial direction $\hat{\bm r}$.  

In Fig.~(\ref{P-for-gen-lambda}a), we observe that as $u_0$ is increased (at fixed $D_\theta$ and $D_t$), the  position of the peak of $P(r)$ shifts away from the center ($r=0$) of the potential, while $P(r) \vert_{r=0}$ becomes a minima. We refer to this shift as the passive-to-active transition. The same transition can also be observed by changing $D_\theta$ at fixed $u_0$ and $D_t$. This transition is described by the curve given by 
\begin{align}
P''\vert_{r=0} \equiv \sum_{n=0}^\infty (n+1) \lambda^{2n} C_{n,0}^{(2n)}(\beta) =0.
\end{align}
In Fig.~(\ref{crossover}a), we plot this curve in the $\lambda -\beta$ plane. We see that at fixed $\beta$, increasing $\lambda$ induces a passive-to-active transition.  However, for fixed $\lambda$ greater than a critical $\lambda^*$, on  increasing $\beta$ from a small value, we  first see a passive-to-active transition. Remarkably, upon further increasing $\beta$, \emph{we observe a re-entrant active-to-passive transition}, as shown in Fig.~(\ref{crossover}b). The first transition was observed in a recent experiment \cite{Takatori_2016_exptAcousticTweezer}, where $\beta$ was varied by changing the trap stiffness $k$.  We here predict a re-entrant transition on increasing $k$ further. In Fig.~(\ref{crossover}c-e), we plot the steady-state distribution of particle positions $\mathbf{r}$ and angles $\theta$ at fixed $\lambda=4$ and varying $\beta$ to represent the three different regions of the phase-diagram in Fig.~(\ref{crossover}a). We observe that while the distribution is peaked at the center of the trap both in the passive phase (c) and the re-entrant phase (e), the distribution of $\theta$ in (e) is akin to that in the active phase (d). Note that the distribution is not peaked at the center of the trap in (d).

Below we  provide a heuristic physical explanation of the two crossovers that we observe.

(i) Passive-to-active crossover --- For a small stiffness $k$ we expect that the particle will effectively behave as a passive particle since it undergoes many orientation changes in the time scale ($\tau_{eq}=(\mu k )^{-1}$) required to reach the passive (Gaussian) steady state. The typical extent of this steady state distribution is 
 $\ell_{th} \sim \sqrt{D_{\rm eff}/(\mu k)}$, where $D_{\rm eff}= D_t+u_0^2/(2 D_\theta)$.
We note that the time taken by particles to reorder their orientation is given by  $\tau_{rot} \sim D_{\theta}^{-1}$, in which time they travel a distance $\ell_{act}=u_0 \tau_{rot}$. As we increase $k$, the length scale $\ell_{th}$ decreases.  We then expect that the passive-like steady state will be stable provided $\ell_{act} \lesssim \ell_{th}$. Otherwise the particles tend to move out and the distribution starts having peaks away from the centre. This gives the passive-to-active crossover and the condition $\ell_{act} \gtrsim \ell_{th}$ for this gives  
\begin{align} 
\beta \gtrsim \frac{2+\lambda^2}{2\lambda^2}. \label{condPtoA} 
\end{align}
Interestingly we note that even for large $\lambda$ one needs a  finite value of $\beta \approx 1/2$ to see the passive-to-active crossover.
 
(ii) Active-to-passive crossover ---  this crossover is observed on further increasing $k$ (or equilvalentsly $\beta$).  Note that  this `reentrant passive phase' is markedly different from the usual passive phase in that, the distribution of the orientation of the particle captures a notrivial signature of activity as explained in Figs.~\eqref{crossover}(c)-(e). In this case we propose the following physical explanation. The activity pushes out the particle  radially so that peaks appear at a distance given by $r_{peak} \sim u_0/\mu k$. If this  distance is small compared to the thermal length scale $\ell'_{th}\sim \sqrt{D_t/(\mu k)}$ (in this case the particles angular motion is restricted and does not contribute to $D_{\rm eff}$), then  we will not see any off-centre peak, even though the particles do mainitain a strong radial  orientation. This then leads to the re-entrant crossover occuring when $\ell'_{th} \gtrsim r_{peak}$ which reduces to
\begin{align}
\beta \gtrsim \lambda^2. \label{condAtoP} 
\end{align}
These  two estimates, Eqs.~(\ref{condPtoA},\ref{condAtoP}), also indicate that 
there exists a critical  $\lambda \approx 1.13$ below which one does not see any of the  crossovers. These heuristic estimates for the crossovers are consistent with our numerical observations in Fig.~\eqref{crossover}a, though a precise quantitative verification would require more work.

Moreover, in the limits $D_{\theta}\to 0$ and $D_{\theta}\to\infty$, we find closed form analytical expressions for this probability distribution. In particular, our approach offers a perturbative solution for the limit $D_t \to 0$, which is non-trivial to obtain for a problem without any translational diffusion ($D_t=0$) \cite{Basu_2018_2D}. We discuss these limits in the next section.

\section{Special limits}  
\label{sec:limits}
\begin{figure}[t]
\includegraphics[width=\linewidth]{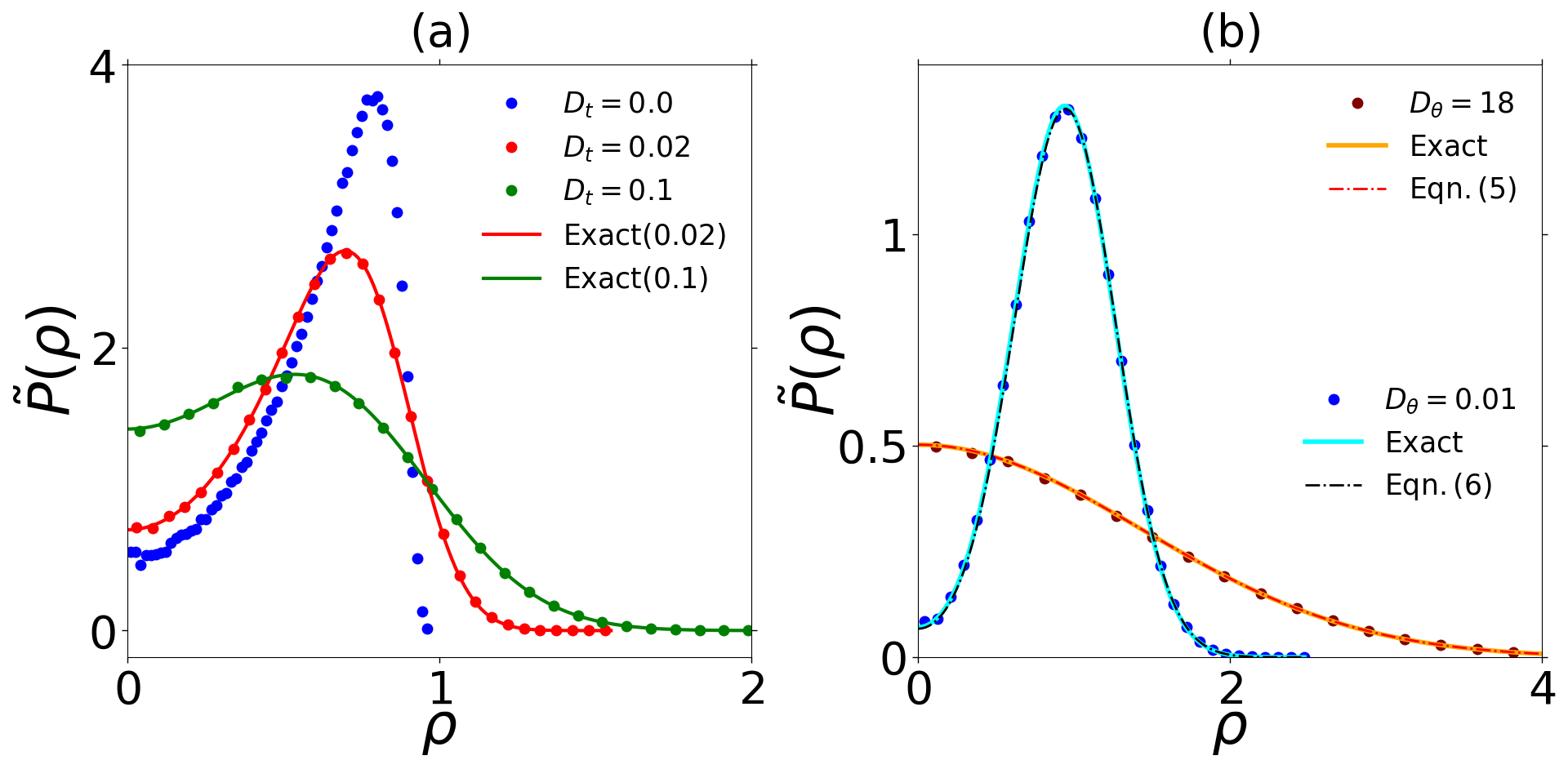}
\caption{
(a) Our analytical result for the radial distribution $\tilde{P}(\rho)=(D_\theta/D_t) P(r)$ with $P(r)$ given by Eq.~\eqref{radPDF} compares well with numerical simulations ($u_0=1, D_{\theta}=1$) even for small values of the translational diffusion $D_t$. (b) The closed form expressions Eq.\eqref{P_Gauaa-Deff} and Eq.\eqref{dth-0-limit}, obtained in the limits $D_{\theta}\to\infty$ and $D_{\theta}\to 0$ respectively, compare very well both with the exact results and the numerical simulations.
For $D_{\theta}=18$, we used $u_0=6$ and $D_t=1$, while for $D_{\theta}=0.01$ we used $u_0=1$ and $D_t=0.1$. In both cases, $\mu=1$ and $k=1$.
}
\label{dth-u0-inf-limit}
\end{figure}

We now discuss certain limiting cases of the radial probability distribution $\tilde{P}(\rho)=(D_\theta/D_t) P(r)$.
We first consider the case with zero translational noise, i.e., $D_t=0$.
Figure~(\ref{dth-u0-inf-limit}a) shows a plot of $\tilde{P}(\rho)$ for $D_t=0$ obtained from our simulations. We observe that $\tilde{P}(\rho)$ has a  finite support  with a peak around $r=1$, in sharp contrast to  the usual Boltzmann distribution. The singular nature of $\tilde{P}(\rho)$ for $D_t=0$ does not allow us to readily obtain an analytical expression in this case.
However, we numerically evaluate our general series expression from Eq.~(\ref{radPDF}) and compare $\tilde{P}(\rho)$ with numerical simulation results for small $D_t$. As shown in  Fig.~(\ref{dth-u0-inf-limit}a), our approach thus allows a systematic way to attain the $D_t\to 0$ limit, although the  computational cost of evaluating Eq.~(\ref{radPDF}) escalates quickly for small values of $D_t$.

We next consider the limit $D_{\theta} \rightarrow \infty $, $u_0 \rightarrow \infty$ such that $\beta \to 0$ with $u_0^2/D_\theta=\lambda^2 D_t$ held constant. In this limit, the stochasticity in $u_0 \hat{{\bf e}} (\theta)$ arising from the rotational diffusion in Eq.~\eqref{eq2} approaches the limit of a zero-mean Gaussian white noise of strength $u_0^2/(2D_\theta)$ \cite{Basu_2018_2D}. This adds to the translational white noise ${\bm \xi}_r(t)$. As a result, we  expect a Gaussian distribution of the form $P(\rho) \simeq  (\mu k)/D_{\rm eff} \exp\big[-\mu k\rho^2/(2D_{\rm eff})\big]$ with an effective diffusion constant $D_{\rm eff}=D_t + {u_0^2}/{(2D_{\theta})}$, as shown in \cite{Basu_2018_2D} for the case $D_t=0$. One can, however, find corrections to this result systematically at different orders in $1/D_\theta$. To accomplish this, we first solve the recursion relations in Eq.~\eqref{recursion-main} perturbatively in powers of $\beta$ to get $C^{(2m)}_{m,0} \simeq (-1/2)^m  \sqrt{{\beta}/{2 \pi}}[1-\beta(7 m^2+m)/8 + O(\beta^{2})]$. Using this in Eq.~(\ref{radPDF}) and performing 
some straightforward calculations we get $\tilde{P}(\rho)$ that takes the form,
\begin{align}
\begin{split}
 \tilde{P}(\rho) &\simeq \frac{\mu k}{D_{\rm{eff}}}~e^{-\frac{\mu k}{2D_{\rm{eff}}}\rho^2}~\left[1 - \frac{\mu k}{D_\theta} \left(  \frac{3}{4} - \frac{5}{4} \frac{\mu k}{D_{\rm{eff}}}\rho^2 \right. \right.  \\
 &~~~~~~~~~~+ \left. \left.  \frac{7}{32} \frac{\mu^2 k^2}{D_{\rm{eff}}^2}\rho^4 \right) + O\left(\frac{1}{D_\theta^2}\right)\right].
 \end{split}
  \label{P_Gauaa-Deff}
\end{align}
in the $D_t\rightarrow 0$ limit (calculation steps and the expression for finite $D_t$ are in  App.~\ref{Appendix-C}). Figure~(\ref{dth-u0-inf-limit}b) validates this asymptotic result with numerical simulations. Note that in the above expression one can take $D_t \to 0$ limit which would correspond to the case of an active Brownian particle without any translational noise.

Finally, we consider the opposite limit $D_\theta \rightarrow 0$, i.e., $\beta \to \infty$, in which the timescale for rotational diffusion of the orientation of $\theta$ diverges as $\sim D_\theta^{-1}$.  
The $\beta \to \infty$ limit can also be obtained by taking $k \to \infty$ limit. 
In this limit, the radial distribution, as shown in \cite{pototsky}, is approximately
\begin{align}
\tilde{P}(\rho) \simeq \frac{1}{Z}\exp\bigg(-\frac{\mu k \rho^2}{2 D_t}\bigg)~I_0\bigg(\frac{u_0 \rho}{D_t}\bigg), \label{dth-0-limit}
\end{align}
where $Z$ is a normalization constant and  $I_0(y)$ is the modified Bessel function of zeroth order. In order to arrive at this result from our general expression in Eq.~\eqref{radPDF}, we look at the asymptotic behavour of the coefficients $C_{n,l}^{(m)}$ for large $\beta$. Note that for large $\beta$, Eq.~\eqref{recursion-main} simplifies to the form $C^{(m)}_{n,l} =  [C^{(m-1)}_{n,l-1} \sqrt{(n+|l|)} - C^{(m-1)}_{n-1,l+1} \sqrt{n}]/{[\sqrt{2 \beta}  (2n + |l|)]}$. It can be easily checked that this equation is satisfied by $C_{n,l}^{(2n+l)} = {(-1)^n} \sqrt{\beta/(2 \pi)} [{n! (n+l)! \beta^{2n + l} 2^{2n + l}}]^{-1/2}$ for $n \geq 0$ and $\l \geq 0$.  Using this result in Eq.~\eqref{radPDF} leads to Eq.~\eqref{dth-0-limit} (see App.~\ref{Appendix-C} for details).

\section{Summary and conclusion}
\label{sec:summ}
To conclude, we have obtained an exact series solution for the non-equilibrium steady state of an ABP, with translational diffusion, confined to a two dimensional harmonic trap. Our analytical results are in good agreement with explicit numerical simulations. We explain a recently observed passive-to-active transition of a harmonically confined ABP with increasing trap-stiffness. Furthermore, we predict a surprising active-to-passive reentrant transition upon a further increase in the confinement strength. The prediction of this reentrant transition is amenable to experimental validation. In fact, using the values from the experimental setup of \cite{Takatori_2016_exptAcousticTweezer},  $D_t\approx 0.1 \mu m^2/s$, $D_\theta^{-1}\approx 25 s$, $u_0\approx 0.25 \mu m/s$, i.e. P\'eclet number $~\lambda \approx 4$, we expect to observe the passive-to-active transition around $\beta_1 \approx 0.35$, i.e., $\mu k_1 \approx 0.01 s^{-1}$ and the re-entrant active-to-passive transition around $\beta_2 \approx 6.8$, i.e., $\mu k_1 \approx 0.27 s^{-1}$. These are plausible parameter values that could be accessed experimentally.
~\\

We thank S. C. Takatori for discussions on the experimental aspects. KM's research is supported by Long Term Visiting Student's Program-2018, ICTS-TIFR. A. Das acknowledges the CEFIPRA postdoctoral fellowship hosted at ICTS-TIFR. AK acknowledges support from DST grant under project No. ECR/2017/000634. KVK's research is supported by the Department of Biotechnology, India, through a Ramalingaswami reentry fellowship and by the Max Planck Society and the Department of Science and Technology, India, through a Max Planck Partner Group at ICTS-TIFR. AK and AD acknowledge the support of the project 5604-2 of the Indo-French Centre for the Promotion of Advanced Research (IFCPAR). We acknowledge support of the Department of Atomic Energy, Government of India, under project no.12-R$\&$D-TFR-5.10-1100.

\bibliography{abd_submit_aps} 

\appendix
\section{Details of calculational of the  steady state distribution}
\label{Appendix-A}
The dimensionless Langevin equation for the active Brownian particle in a two dimensional harmonic trap is given by,
\begin{align}\begin{split}
\frac{d{\bf r}}{dt} &= \lambda~ \hat{{\bf e}} (\theta) - \beta {\bf r} + \sqrt{2} \xi_r(t),\\ 
\frac{d \theta}{dt} &= \sqrt{2} \xi_{\theta}(t),\end{split}\label{eq4}
\end{align}
where the dimensionless parameters $\lambda$ and $\beta$ are defined as $\lambda = \frac{u_0}{\sqrt{D_{\theta} D_t}}, \beta = \frac{\mu k}{D_{\theta} }.$
The corresponding Fokker-Planck equation is
\begin{eqnarray}
\frac{\partial \mathcal{P}}{\partial t} = -\nabla \cdot (\lambda \hat{e} -\beta {\bf r} -\nabla)P + \frac{\partial^2 \mathcal{P}}{\partial \theta^2} .\label{Aeq6}
\end{eqnarray}
In the two dimensional polar coordinates the above equation takes the form
\begin{eqnarray}
 \frac{\partial \mathcal{P}}{\partial t}&=& \lambda \bigg[\cos \chi \frac{\partial \mathcal{P}}{\partial r} + \frac{\sin \chi}{r} \frac{\partial \mathcal{P}}{\partial \varphi}\bigg] \nonumber\\
&~&+ \frac{1}{r} \frac{\partial}{\partial r} \bigg[ r \bigg( \frac{\partial}{\partial r} + \beta r \bigg) \mathcal{P} \bigg] + \frac{1}{r^2} \frac{\partial^2 \mathcal{P}}{\partial \varphi^2} + \frac{\partial^2 \mathcal{P}}{\partial \theta^2},\nonumber
\end{eqnarray}
where $\chi=\theta-\varphi$.
As discussed in the main text, in the steady state given by ${\partial \mathcal{P}}/{\partial t}=0$, we expect the solution to have azimuthal symmetry, $\mathcal{P}(r,\varphi,\theta)=\mathcal{P}(r,\chi)$.  Hence, in the steady state, after making the replacements $\frac{\partial}{\partial \theta}=\frac{\partial}{\partial \chi}$, $\frac{\partial}{\partial \varphi}=-\frac{\partial}{\partial \chi}$, we find that the equation satisfied by the steady state probability distribution $\mathcal{P}(r,\chi)$ can be written in the form (Eq.~\eqref{eq14} of main text)
\begin{align}
\mathcal{L}_0 \mathcal{P} &= \lambda \, \mathcal{L}_1 \mathcal{P}, \label{Aeq14}
\end{align}
where 
\begin{align}
\mathcal{L}_0 &=  \frac{1}{r} \frac{\partial}{\partial r} \bigg[ r \bigg( \frac{\partial}{\partial r} + \beta r \bigg) \bigg]
+ \frac{1}{r^2} \frac{\partial^2 }{\partial \chi^2} + \frac{\partial^2 }{\partial \chi^2}~, \label{AL0} \\
 \mathcal{L}_1 &= \cos \chi \frac{\partial}{\partial r} - \frac{\sin \chi}{r} \frac{\partial}{\partial \chi}~. \label{AL1}
\end{align}
We try a  series solution of  the  form 
\begin{align}
\mathcal{P}(r, \chi) = \sum_{m=0}^\infty \lambda^{m} \mathcal{P}^{(m)}(r,\chi). \label{AP-joint}
\end{align}
Inserting this in Eq.~\eqref{Aeq14} and equating powers of $\lambda$ on both sides, we get 
\begin{eqnarray}
\mathcal{L}_0 \mathcal{P}^{(0)} = 0,~~\mathcal{L}_0 \mathcal{P}^{(m)} = \mathcal{L}_1 \mathcal{P}^{(m-1)}~, \quad m=1,2,\ldots\; \label{seriessol}
\end{eqnarray}
The first equation ($m=0$) above can be solved to give $\mathcal{P}^{(0)} (r, \chi) = e^{-\beta r^2/2}/{Z}$, where $Z=2 \pi \int_0^\infty dr \, r e^{-\beta r^2/2}$. 
As mentioned in the main text, if the eigenfunctions and eigenvalues of $\mathcal{L}_0$ are known explicitly and if they form a complete basis, then at any order $m$, the term on the right-hand-side of Eq.~(\ref{seriessol}) calculated from the previous order ($m-1$) solution can be expanded in the eigenbasis of $\mathcal{L}_0$ and one obtains $\mathcal{P}^{(m)}(r,\chi)$. The procedure is detailed in the following.\\

\begin{figure}[t]
\includegraphics[width=0.9\linewidth]{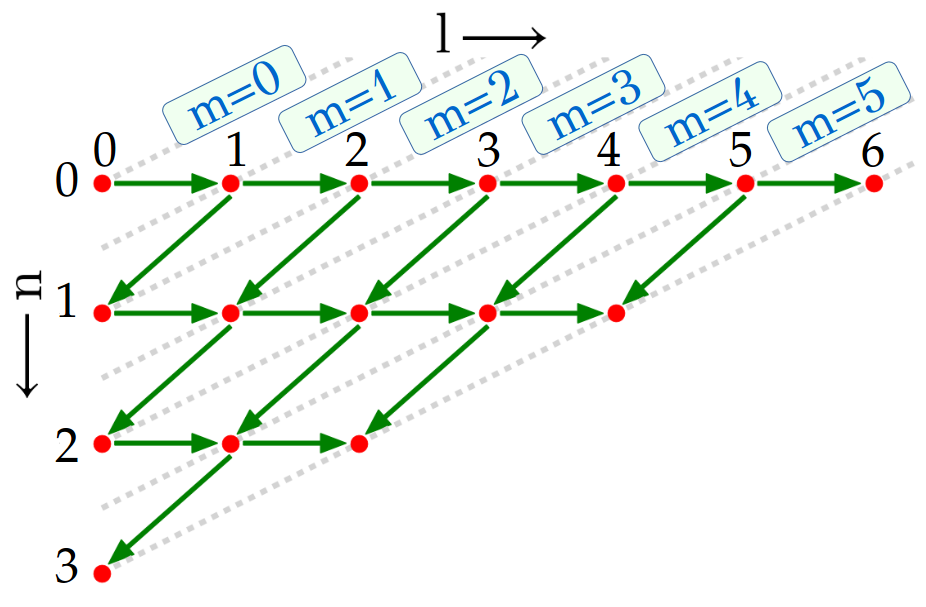}
\caption{Schematic illustrating the structure of the recursion relations in Eqs.~\eqref{Arecursion-main} in the main text. Each $C_{n, l}^{(m)}$ gets contributions from the coefficient at the previous order $m-1$  both from $(n-1, l+1)$ and $(n,l-1)$. The order $m$ ($=2n+l$) is constant on the dashed lines.}
\label{fig:recurstion_table}
\end{figure}

\emph{Eigensystem of $\mathcal{L}_0$}:
Let $\phi(r,\chi)$ be the eigenvector of $\mathcal{L}_0$ corresponding to the eigenvalue $\nu$ satisfying the equation $\mathcal{L}_0 \phi = \nu \phi$.  With the transformation $\phi= e^{-\beta V(r)/2} \, \psi$,  we find that $\psi$ satisfies the equation $H \psi = \nu \psi$ where $H= e^{-\beta V(r)/2} \mathcal{L}_0 e^{\beta V(r)/2}$ is a Hermitian Schr\"odinger operator. Explicitly, the eigenvalue equation is
\begin{eqnarray}
\bigg[\frac{1}{r} \frac{\partial}{\partial r} \bigg(r  \frac{\partial}{\partial r}\bigg) + \frac{1}{r^2}  
\frac{\partial^2}{\partial \chi^2} +  \beta  - \frac{\beta^2 r^2}{4}  + \frac{\partial^2}{\partial \chi^2} \bigg] \psi = \nu \psi.\quad~~~ \label{eq32}
\end{eqnarray}
The above equation without the last $\partial^2/\partial \chi^2$ term corresponds to the Schr\"{o}dinger equation of an isotropic two-dimensional harmonic oscillator, for which the eigenfunctions ($\psi_{n,l}$) and eigenvalues ($\Lambda_{n,l}$)  are known exactly~\cite{Yanez_1994_polarHO}:
\begin{eqnarray}
\psi_{n,l}(r, \chi) &=& \bigg[ \frac{n! (\frac{\beta}{2})^{|l|+1}}{\pi \Gamma(n+|l|+1)} 
\bigg]^{\frac{1}{2}} r^{|l|} e^{-\frac{\beta r^2}{4}} L_n^{|l|} \bigg(\frac{\beta r^2}{2}\bigg)  e^{il \chi}~, \nonumber
\end{eqnarray}
with $\Lambda_{n,l} = -\beta  (2n + |l|) ~$ where $n \ge 0,~l$ are integers, $\Gamma$ is the gamma function, and $L_n^{|l|}(x)$ is the generalized Laguerre polynomial. Note that the presence of the last  $\partial^2/\partial \chi^2$ term in Eq.~\eqref{eq32} does not change the eigenfunctions while the eigenvalues get modified to
\begin{eqnarray}
\nu_{n,l} &=& \Lambda_{n,l}-l^2=-\beta  (2n + |l|)-l^2 ~. \label{eq38}
\end{eqnarray}
The $\psi_{n,l}$ form a set of basis functions. Since the Fokker-Planck operator $\mathcal{L}_0$ is not a Hermitian operator, its right and left eigenfunctions are different.  
The right eigenfunctions are  given by $$\phi_{n,l}=e^{-\beta r^2/4} \psi_{n,l}.$$ It can be shown that the left eigenfunctions are given by $$\tilde{\phi}_{n,l}=e^{\beta r^2/4} \psi^*_{n,l},$$ and these satisfy the orthogonality condition $$\int_0^\infty dr \, r\int_0^{2 \pi} d \chi ~\tilde{\phi}_{n,l} \phi_{n',l'}=\delta_{n, n'} \delta_{l, l'}.$$  
Next, we expand the solutions, at different order, in the $\mathcal{L}_0$-basis: 
\begin{equation}
\mathcal{P}^{(m)}(r, \chi) = \sum_{n,l} C_{n, l}^{(m)} \phi_{n,l}(r, \chi). \label{p^m}
\end{equation}
 Inserting this in Eq.~(\ref{seriessol}), and using the orthogonality of the basis states, we find
 $C_{n,l}^{(m)}=({1}/{\nu_{n,l}}) \int_0^\infty dr \, r\int_0^{2 \pi} d \chi~ \mathcal{L}_1 \mathcal{P}^{(m-1)} \tilde{\phi}_{n,l}$. Using the specific form of $\mathcal{L}_1$ given in Eq.~\eqref{AL1}, we expand  $\mathcal{L}_1 \mathcal{P}^{(m-1)} \equiv f^{(m)} (r, \chi)$ in the same basis set and, 
on using the above mentioned orthonormality, we find that the coefficients satisfy the recursion relations  ( Eqs.~\eqref{recursion-main} in the main text)
\begin{align}
C^{(m)}_{n,l} &=  \frac{C^{(m-1)}_{n,l-1} \sqrt{(n+|l|) \frac{\beta}{2}} - C^{(m-1)}_{n-1,l+1} 
\sqrt{n \frac{\beta }{2}} }{\beta  (2n + |l|) + l^2}~,~l>0 \nonumber \\
C^{(m)}_{n,0} &= -{C^{(m-1)}_{n-1,1}}/{\sqrt{2 \beta n}}~.  
\label{Arecursion-main}
\end{align}
with $C^{(0)}_{0,0}=1/\sqrt{2 \beta}$. Thus we have 
\begin{align} \label{Pmr}
\mathcal{P}^{(m)}(r,\chi)&=
\sum_{n,l}C^{(m)}_{n,l} e^{-\frac{\beta r^2}{4}} \psi_{n,l}(r,\chi),~m=2n+|l|,~n\ge 0,\\
f^{(m)} (r, \chi) &= \sum B^{(m)}_{n,l} e^{-\frac{\beta r^2}{4}} \psi_{n,l}~.
\end{align}
Inserting these expansions into Eq.~(\ref{seriessol}) and using the orthonormality of $\psi_{n,l}$, we get
\begin{eqnarray}
C^{(m)}_{n,l} &=& \frac{B^{(m)}_{n,l}}{\nu_{n,l}} = \frac{B^{(m)}_{n,l}}{- \beta (2n+l) - l^2}, \\
{\rm where}~~B^{(m)}_{n,l} &=& \int\int rdr d\chi~ f^{(m)}(r,\chi) e^{\beta r^2/4} \psi_{n,l}~.\nonumber
\end{eqnarray}
Now, using the explicit forms of the eigenfunctions
 \begin{eqnarray}
\psi_{n,l}(r, \chi) &=& \bigg[ \frac{n! (\frac{\beta}{2})^{|l|+1}}{\pi \Gamma(n+|l|+1)} 
\bigg]^{\frac{1}{2}} r^{|l|} e^{-\frac{\beta r^2}{4}} L_n^{|l|} \bigg(\frac{\beta r^2}{2}\bigg)  e^{il \chi}~, \nonumber
\end{eqnarray}
and defining $ N^{(m)}_{n,l}=C^{(m)}_{n,l} \sqrt{\frac{2n!  \big( \frac{\beta }{2} \big)^{|l|+1}}{2 \pi (n+|l|)!}}$,
we get
\begin{eqnarray}
\mathcal{P}^{(m)} &=& \sum_{n,l} C^{(m)}_{n,l} e^{-\frac{\beta r^2}{4}} \psi_{n,l} 
 = \sum_{n,l} N^{(m)}_{n,l} r^{|l|} e^{il \chi } e^{-\frac{\beta r^2}{2}} L_n^{|l|} \bigg( \frac{\beta r^2}{2} \bigg),\nonumber
\end{eqnarray}
and the source term
\begin{widetext}
\begin{eqnarray}
&&f^{(m)} (r, \chi)= \mathcal{L}_1 \mathcal{P}^{(m)} = \cos \chi \frac{\partial \mathcal{P}^{(m)}}{\partial r} -
\frac{\sin \chi}{r} \frac{\partial \mathcal{P}^{(m)}}{\partial \chi}\nonumber\\
&~&= \sum_{n,l} \frac{N^{(m)}_{n,l}}{2} \bigg[ r^{|l|-1} L_n^l(x) \bigg( |l| \big[ e^{i(l+1) \chi} + e^{i (l-1) \chi} \big] - l \big[ e^{i(l+1) \chi}   - e^{i (l-1) \chi} \big] \bigg) - \beta r^{|l|+1} L_n^{|l|+1}(x) \big( e^{i(l+1) \chi} + e^{i (l-1) \chi} \big)  \bigg]  e^{-\frac{\beta r^2}{2}}, \nonumber
\end{eqnarray}
\end{widetext}
where $x=\beta r^2/2$. Using the orthonormality of the Laguerre polynomials and the azimuthal eigenfunctions, we finally obtain the following recursion relation
\begin{equation}
C^{(m)}_{n,l} =  \frac{C^{(m-1)}_{n,l-1} \sqrt{(n+l) \frac{\beta}{2}} - C^{(m-1)}_{n-1,l+1} \sqrt{n \frac{\beta}{2}} }{\beta (2n + l) + l^2},~~ n > 0,~l>0~
\end{equation}
and from radial symmetry,
\begin{equation}
C^{(m)}_{n,-l} = C^{(m)}_{n,l}~. 
\end{equation}
The `boundary' recursion equations are given by
\begin{eqnarray}
C_{0, l}^{(m)} &=&  \frac{C_{0,l-1}^{(m-1)} \sqrt{l \frac{\beta}{2}}}{\beta l + l^2}, ~~l>0, \\
C_{n, 0}^{(m)} &=& - \frac{C_{n-1,1}^{(m-1)} \sqrt{n \frac{\beta}{2}}}{\beta n } ,~n>0~.
\end{eqnarray}
The recursion relations can  be solved iteratively, starting from the ``initial'' condition
\begin{eqnarray}
C^{(0)}_{0, 0} = \sqrt{ \frac{\beta}{2 \pi }},
\end{eqnarray}
and the boundary condition, $C^{(m)}_{-1,l}=0,~l\ge 0.$
%
The structure of the recursion relations is illustrated in Fig.~\eqref{fig:recurstion_table}. Here the red filled discs represent $C^{(m)}_{n,l}$ with the horizontal and the vertical axes representing $l$ and $n$ respectively, while the dashed lines correspond to constant $m=2n+l$. 
The coefficients $C_{n,l}^{(m)}$, at a given $m$,  can be obtained from the coefficients $C_{n,l-1}^{(m-1)}$ and $C_{n-1,l+1}^{(m-1)}$ on the $(m-1)$ line, as indicated by the two   arrows meeting at the point $(n,l)$. One can think of the index $m$ as a time index and $n,l$ as spatial indices. Thus, starting from the initial condition localized at $n=0,~l=0$,  these recursions prescribe a time evolution in the $m$ direction. 
Following this procedure, we can in principle find all the coefficients for arbitrary $n,l$, and hence $\mathcal{P}^{(m)}$ at any order $m$.
So, starting with $C^{(0)}_{0, 0}$, all other $C^{(m)}_{n,l}$ can be obtained using the above recursion relations, and hence we get an explicit series solution for $\mathcal{P}(r, \chi)$ given by Eqs.~(\ref{AP-joint},~\ref{Pmr}). In the following we list first few coefficients,
\begin{eqnarray*}
&& C^{(0)}_{0,0} = \sqrt{\frac{\beta}{2 \pi}},~ C^{(1)}_{0,1} = \frac{\beta }{\sqrt{\pi } (2 \beta +2)},~ C^{(2)}_{0,2} = \frac{(\beta )^{3/2}}{4 \sqrt{\pi } (\beta +1) (\beta +2)},\\
&&~ C^{(2)}_{1,0} = -\frac{\sqrt{\beta }}{\sqrt{2 \pi } (2 \beta +2)},~ C^{(3)}_{0,3} = \frac{ \beta ^2}{4 \sqrt{6 \pi } (\beta +1) (\beta +2) (\beta +3)},\\
&& C^{(3)}_{1,1} = -\frac{\beta  (3 \beta +4)}{4 \sqrt{2 \pi } (\beta +1) (\beta +2) (3 \beta +1)}
\end{eqnarray*}
and so on. Integrating over $\chi$ we can obtain the radial distribution $P(r)$. It is easy to see that only the $l=0$ terms contribute to the radial distribution which gives,
\begin{eqnarray}
P(r) &=& \int d\chi \sum_{n=0}^{\infty} \lambda^{2n} C^{(2 n)}_{n, 0} e^{-\frac{\beta r^2}{4}} \psi_{n,0} \nonumber \\
&=& \int d\chi  \sum_{n=0}^{\infty} \lambda^{2n} C^{(2 n)}_{n, 0} e^{-\frac{\beta r^2}{2}} 
\bigg[\frac{\beta }{2 \pi} \bigg]^{\frac{1}{2}} L_{n}\bigg(\frac{\beta r^2}{2}\bigg),~~~~~~
\end{eqnarray}
where $L_{n}$ is Laguerre polynomial of order $n$. Simplifying the above, we get,
\[
\boxed{ P(r) = \sum_{n=0}^{\infty} \lambda^{2n} \sqrt{2 \pi \beta} C^{(2n)}_{{n}, 0} e^{-\frac{\beta r^2}{2}} L_{{n}} \bigg( \frac{\beta r^2}{2} \bigg) }
\]
which is Eq.~\eqref{radPDF} of the main article. In the next section we demonstrate the results for first few $\mathcal{P}^{(m)}(r,\chi)$ calculated in this method.

\section{Explicit expressions for  $\mathcal{P}^{(m)}(r,\chi)$ for $m=1,2,3,4$}\label{Appendix-B}
\paragraph{Zeroth order} ($\mathcal{P}^{(0)}$):
\begin{eqnarray}
\mathcal{P}^{(0)} (r, \chi) = \frac{\beta }{2 \pi} e^{- \frac{\beta r^2}{2}}
\end{eqnarray}
This is the equilibrium distribution (for $\lambda = 0$).\\

\paragraph{First order} ($\mathcal{P}^{(1)}$):
\begin{eqnarray}
f^{(1)} (r, \chi) &=& \mathcal{L}_{1} \mathcal{P}^{(0)} = -\frac{\beta ^2 r \cos (\chi ) e^{-\frac{1}{2} \beta r^2}}{2 \pi },\nonumber \\
\mathcal{P}^{(1)} (r, \chi) &=& \frac{\beta ^2  r \cos (\chi ) e^{-\frac{1}{2} \beta r^2}}{2 \pi  (\beta +1 )}~.
\end{eqnarray}
\begin{widetext}
\paragraph{Second order} ($\mathcal{P}^{(2)}$):
\begin{eqnarray}
 f^{(2)} (r, \chi) &=& \mathcal{L}_{1} \mathcal{P}^{(1)} = -\frac{\beta ^2 e^{-\frac{1}{2} \beta r^2} (\beta r^2 \cos (2 \chi )+\beta
   r^2-2)}{4 \pi ( \beta+1 )},\nonumber\\
\mathcal{P}^{(2)} (r, \chi) &=& \frac{\beta e^{-\frac{1}{2} \beta r^2} \big(\beta ^2 r^2 \cos (2 \chi )+(\beta +2)
   (\beta r^2-2)\big)}{8 \pi  (\beta+1) (\beta +2)}
\end{eqnarray}

\paragraph{Third order} ($\mathcal{P}^{(3)}$):
\begin{eqnarray}
 f^{(3)} (r, \chi) &=& \mathcal{L}_{1} \mathcal{P}^{(2)} = -\frac{\beta ^2 r \cos (\chi ) e^{-\frac{1}{2} \beta r^2} \big(\beta ^2 r^2 \cos (2
   \chi )+\beta (r^2 (\beta +2)-6)-8\big)}{8 \pi  (\beta +1) (\beta +2)}, \nonumber \\
 \mathcal{P}^{(3)} (r, \chi) &=& \frac{\beta ^2 r e^{-\frac{1}{2} \beta r^2} \big(\beta ^2 r^2 (3 \beta+1) \cos (3
   \chi )+3 (\beta+3) (3 \beta +4) \cos (\chi ) (\beta r^2-4)\big)}{48 \pi 
   (\beta +1) (\beta+2) (\beta +3) (3 \beta +1)}
\end{eqnarray}

\paragraph{Fourth order} ($\mathcal{P}^{(4)}$):
\begin{eqnarray}
\mathcal{P}^{(4)} (r, \chi) = \frac{\beta e^{-\frac{1}{2} \beta  r^2}}{768 \pi  (\beta +1)^2 (\beta +2) (\beta +3) (\beta +4) (3 \beta +1)} ~
     \bigg(\beta ^4 r^4 (\beta +1) (3 \beta +1) \cos (4 \chi )\nonumber\\
    +4 \beta ^2 r^2 (\beta +4) (3 \beta^2 +10 \beta +9) \cos (2 \chi ) (\beta r^2-6) \nonumber \\
    +6 (\beta +1) (\beta +3) (\beta +4) (3 \beta +4) \big(\beta r^2 (\beta  r^2-8)+8\big)\bigg)~.
\end{eqnarray}
\end{widetext}

\section{Special limiting cases}\label{Appendix-C}
With the replacements, $\beta = \frac{\mu k }{D_{\theta}},~r = \rho \sqrt{\frac{D_{\theta}}{D_{t}}},~\lambda = \frac{u_0}{\sqrt{D_{\theta} D_{t}}},~ t = \tau D_{\theta}$,
the probability distribution $\tilde{P}(\rho)$ in the dimensionful variables will be,
\begin{eqnarray}
\tilde{P}(\rho) = P(r) \frac{r}{\rho} \bigg\vert \frac{\partial r}{\partial \rho} \bigg\vert  = P(r) \frac{D_{\theta}}{D_t},\nonumber
\end{eqnarray}
i.e., in equilibrium when activity is absent, $\tilde{P}^{(0)} (\rho) = \frac{\mu k }{D_{t}} e^{-\frac{\mu k}{2 D_{t}} \rho^2}$,
and in the non-equilibrium steady state in presence of nonzero activity,
\begin{eqnarray}
\tilde{P}(\rho) && = \frac{D_{\theta}}{D_t} \sum_{m=0}^{\infty} \bigg( \frac{u_0}{\sqrt{D_{\theta} D_{t} }} \bigg)^{2m} \sqrt{\frac{2 \pi \mu k}{D_{\theta}}} ~ C^{(2m)}_{m,0} \nonumber\\
&&~~~~~~~~~~~~~~~~~~~~~~~~\times e^{-\frac{\mu k \rho^2}{2 D_t}} L_m \bigg( \frac{\mu k \rho^2}{2 D_{t}} \bigg)
\end{eqnarray}

\subsection{Limit $D_{\theta} \rightarrow \infty $, $u_0 \rightarrow \infty$, i.e. $\beta \rightarrow 0$ and $\lambda =$  finite}

In the limit $\beta \rightarrow 0$, the coefficients $C^{(m)}_{n,l}$ (see Fig.~\eqref{fig:recurstion_table}) take the form :

\begin{align*}
\begin{array}{c|ccccc}
n \setminus l & 0 & 1 & 2 & 3 & 4 \\ 
\hline 
0 & \sqrt{\frac{\beta}{2 \pi}} & \frac{\beta}{2 \sqrt{\pi}} & \frac{\beta^{3/2}}{8\sqrt{\pi}} & \frac{\beta^2}{24\sqrt{6 \pi}} &
\frac{\beta^{5/2}}{384 \sqrt{3\pi}} \\ 
1 & - \frac{1}{2} \sqrt{\frac{\beta}{2 \pi}} & - \frac{\beta}{2 \sqrt{2 \pi}} & \frac{3\beta^{3/2}}{16\sqrt{3\pi}} &    &    \\ 
2 & \frac{1}{4} \sqrt{\frac{\beta}{2 \pi}} &    &    &    &   
\end{array} 
\end{align*}
~\\
to leading order in $\beta$. It is evident from the table that the $l=0$ terms are proportional to $\sqrt{\beta}$, 
and the higher $l$ terms come with higher powers of $\beta$. Our aim is to evaluate $C^{(2m)}_{m,0}$, which, from Eq.~\eqref{Arecursion-main} in the main text, depends on $C^{(2m-1)}_{m-1,l}$, that in turn depends on other $n~{\rm and}~l$ terms.
However since the terms with $l$=2 and above appears with higher powers in $\beta$, to obtain the leading contribution in $\beta \rightarrow 0$ limit, we truncate the recursion equation at $l=1$,
\begin{eqnarray}
&& C^{(m)}_{n,1} \approx \frac{\sqrt{\beta (n+1)}}{\sqrt{2}} C^{(m-1)}_{n,0},~\mbox{and}\\
&& C^{(2m)}_{m,0} = - \frac{C^{(2m-1)}_{m-1,1}}{\sqrt{2 \beta m}}
\end{eqnarray}
Combining these two, we obtain,
\begin{eqnarray}
C^{(2m)}_{m,0} &=& \bigg( - \frac{1}{2} \bigg)^{m} \sqrt{\frac{\beta}{2 \pi}} .
\end{eqnarray}
Using the above in the series solution in dimensionful quantities, the steady state probability distribution takes the form,
\begin{eqnarray}
\tilde{P}(\rho) &=& \frac{D_{\theta}}{D_t} \sum_{m=0}^{\infty} \lambda^{2m} \sqrt{2 \pi \beta} C^{(2m)}_{m,0} e^{-\frac{\mu k}{2 D_{t}} \rho^2} L_{m}\bigg(\frac{\mu k}{2 D_{t}} \rho^2 \bigg) \nonumber \\
     &=& \frac{\mu k}{D_t} e^{-\frac{\mu k}{2 D_{t}} \rho^2} \sum_{m=0}^{\infty} \bigg( - \frac{\lambda^{2}}{2} \bigg)^{m} L_{m}\bigg(\frac{\mu k}{2 D_{t}} \rho^2\bigg) \nonumber\\
     &=& \frac{\mu k}{D_t + \frac{u_0^2}{2D_{\theta}}} \exp\bigg[- \frac{1}{2} \frac{\mu k \rho^2}{ D_t + \frac{u_0^2}{2D_{\theta}}} \bigg]
\end{eqnarray}
where we have replaced the sum using generating function of Laguerre polynomials, $\sum_{n=0}^{\infty} L_n(x)t^n = \frac{1}{1-t} \exp[-\frac{tx}{1-t}]$, with $t = -\frac{\lambda^2}{2}$ and $x = \frac{\mu k \rho^2}{2D_t}$.
For $D_t \rightarrow 0$ , this reduces to the expression, $\tilde{P}(\rho) =  \frac{2 \mu k D_{\theta}}{u_0^2} 
\exp[- \frac{2 \mu k D_{\theta} \rho^2}{ u_0^2} ]$ obtained in \cite{Basu_2018_2D}.\\
~\\
\paragraph*{Solution to $\mathcal{O}(\beta)$:} To find the leading correction to $\mathcal{O}(\beta)$, we keep one more order in the recursion for the $C^{(m)}_{n,l}$. The relevant recursion relations truncated at $l=2$ are,
\begin{eqnarray}
&& C^{(2n)}_{n,0} = -\frac{1}{\sqrt{2\beta n}}C^{(2n-1)}_{n-1,1} ~\mbox{(boundary condition)} \label{recursion_bc}\\
&& C^{(2n+1)}_{n,1} = \sqrt{\frac{\beta}{2}}~\frac{C^{(2n)}_{n,0}\sqrt{n+1}-C^{(2n)}_{n-1,2}\sqrt{n}}{1+\beta (2n+1)} \label{recursion}\\
&& C^{(2n+2)}_{n,2} \approx \frac{1}{2}\sqrt{\frac{\beta}{2}}~\frac{\sqrt{n+2}}{2+\beta(n+1)}C^{(2n+1)}_{n,1}\label{truncated}
\end{eqnarray}
~\\
The above set of equations can be simplified as follows.
Replacing $n$ by $n-1$ in Eq.(\ref{recursion}), we get, $C^{(2n-1)}_{n-1,1} = \sqrt{\frac{\beta}{2}}~\frac{C^{(2n-2)}_{n-1,0}\sqrt{n}-C^{(2n-2)}_{n-2,2}\sqrt{n-1}}{1+\beta (2n-1)}$.
$C^{(2n-2)}_{n-2,2}$ is obtained from Eq.(\ref{truncated}) as, $C^{(2n-2)}_{n-2,2} = \frac{1}{2}\sqrt{\frac{\beta}{2}}~\frac{\sqrt{n}}{2+\beta(n-1)}C^{(2n-3)}_{n-2,1}$, which, using Eq.(\ref{recursion_bc}), gives, $C^{(2n-2)}_{n-2,2} = \frac{\beta}{2}\frac{\sqrt{n(n-1)}}{2+\beta(n-1)}C^{(2n-2)}_{n-1,0}$. Inserting the last expression in
the equation for $C^{(2n-1)}_{n-1,1}$ and applying Eq.(\ref{recursion_bc}) once again, we finally obtain a simplified relation between the coefficients in the radial distribution function,
\begin{eqnarray}
 C^{(2n)}_{n,0} &=& -\frac{1}{4}~\frac{4+3\beta(n-1)}{(2+\beta(n-1))(1+\beta(2n-1))}C^{(2n-2)}_{n-1,0}\nonumber\\
 &&\approx \bigg[-\bigg(\frac{1}{2}+\frac{3\beta}{8}\bigg)+
 \frac{7\beta}{8}n\bigg]C^{(2n-2)}_{n-1,0},~n\ge 1. ~~~\label{recursion_l2}
\end{eqnarray}
Its solution correct to $\mathcal{O}(\beta)$ is,
\begin{equation}
 C^{(2n)}_{n,0}\approx \bigg(-\frac{1}{2}\bigg)^n\bigg[1-\frac{\beta}{8}~(7n^2+n) \bigg]C^{(0)}_{0,0},
\end{equation}
~\\
with $C^{(0)}_{0,0}=\sqrt{\frac{\beta}{2\pi}}$. Therefore, the radial distribution function takes the form,
$$P(r) = \beta e^{-\beta r^2/2}\sum_{n=0}^{\infty} \bigg(-\frac{\lambda^2}{2}\bigg)^n\bigg[1-\frac{\beta}{8}~(7n^2+n) \bigg]L_n(\beta r^2/2)$$
In the limit $D_{\theta} \rightarrow \infty$, i.e., $\beta \rightarrow 0$, the quantity $y^2 = \beta r^2$ remains finite. So the correct 
radial distribution to $\mathcal{O}(\beta)$ in terms of the dimensionless quantities is,
\begin{eqnarray}
 P(y) &=& e^{-y^2/2}\sum_{n=0}^{\infty} \bigg(-\frac{\lambda^2}{2}\bigg)^n\bigg[1-\frac{\beta}{8}~(7n^2+n) \bigg]L_n(y^2/2).\nonumber\\\label{series}
\end{eqnarray}
Using again the generating function of Laguerre polynomials, $G_x(t)=\sum_{n=0}^{\infty} t^n L_n(x) = \frac{1}{1-t}\exp(-\frac{tx}{1-t})$, we evaluate the terms in the sum above to get: 
$$\sum_{n=0}^{\infty} n t^n L_n(x) = t \frac{dG}{dt} = t \big( \frac{1}{1-t} - \frac{x}{(1-t)^2} \big) G,$$
and 
\begin{align} &\sum_{n=0}^{\infty} n^2 t^n L_n(x) = t^2 \frac{d^2G}{dt^2} + t \frac{dG}{dt} \nonumber \\&= \bigg[ t^2\big( \frac{2}{(1-t)^2} - \frac{4x}{(1-t)^3} + 
\frac{x^2}{(1-t)^4} \big) + t \big( \frac{1}{1-t} - \frac{x}{(1-t)^2} \big)\bigg]G.  \nonumber
\end{align}
Inserting these in Eq.(\ref{series}), we get
\begin{widetext}
\begin{eqnarray}
 P(y) &=& \frac{e^{-\frac{y^2/2}{1+\lambda^2/2}}}{1+\lambda^2/2}\left[\bigg\lbrace 1+\beta\frac{\lambda^2}{2+\lambda^2}-\frac{7\beta}{4}
 \bigg( \frac{\lambda^2}{2+\lambda^2} \bigg)^2 \bigg\rbrace 
 + \frac{\beta \lambda^2}{(2+\lambda^2)^2} \bigg\lbrace -1 +
 \frac{7\lambda^2}{2(2+\lambda^2)} \bigg\rbrace y^2 -\frac{7\beta\lambda^4}{8(2+\lambda^2)^4}y^4 \right]. \label{dimensionless}
\end{eqnarray}
Reverting to dimensionful quantities,
\begin{eqnarray}
 \tilde{P}(\rho)&=& \frac{\mu k}{D_{\rm{eff}}}~e^{-\frac{\mu k}{2D_{\rm{eff}}}\rho^2}~\bigg[1 + \frac{\mu k}{D_{\theta}}\bigg(1-\frac{D_t}{D_{\rm{eff}}}\bigg)
\bigg\lbrace \bigg(1 - \frac{7}{4}\bigg(1-\frac{D_t}{D_{\rm{eff}}}\bigg)\bigg) \nonumber\\
&&~~~~~~~~~~~~~~~~~~~~~~~~~  + \frac{1}{2} \bigg(-1 + \frac{7}{2}\bigg(1-\frac{D_t}{D_{\rm{eff}}}\bigg)\bigg)\frac{\mu k}{D_{\rm{eff}}} \rho^2 - 
\frac{7}{32} \bigg(1-\frac{D_t}{D_{\rm{eff}}}\bigg) \frac{\mu^2k^2}{D_{\rm{eff}}^2} \rho^4 \bigg\rbrace \bigg], \label{dimensionful-small-beta}
\end{eqnarray}
~\\
where $D_{\rm{eff}}=D_t+\frac{v^2}{2D_{\theta}}$. In the limit $D_t \rightarrow 0 ~(\mbox{i.e.}~
\lambda \rightarrow \infty)$, the radial distribution function takes the form,
\begin{eqnarray}
 \tilde{P}(\rho)
  &&= \frac{\mu k}{D_{\rm{eff}}}~e^{-\frac{\mu k}{2D_{\rm{eff}}}\rho^2}~\bigg[1 + \frac{\mu k}{D_\theta} \bigg( - \frac{3}{4} + \frac{5}{4} \frac{\mu k}{D_{\rm{eff}}}\rho^2 - \frac{7}{32} \frac{\mu^2 k^2}{D_{\rm{eff}}^2}\rho^4 \bigg) \bigg]\label{dimensionful-small-beta-Dt0}
\end{eqnarray}
which is Eq.~\eqref{P_Gauaa-Deff} of the main text.
\end{widetext}

\subsection{Limit $D_\theta \rightarrow 0~ (\beta \rightarrow \infty)$}
In this limit, the recursion relation takes the form

\begin{eqnarray}
C^{(m)}_{n,l} \xrightarrow{\beta \rightarrow \infty} \frac{\sqrt{n+l} C^{(m-1)}_{n,l-1}  - \sqrt{n} C^{(m-1)}_{n-1,l+1}}{(2n+1)\sqrt{2 \beta}}, \label{recursionInfty}
\end{eqnarray}
\begin{widetext}
solving which we get the following table for  the  coefficients 
\begin{align*}
\begin{array}{c|ccccccc}
n \setminus l  & 0 & 1 & 2 & 3 & 4 & 5 & 6 \\ 
\hline
0 & \sqrt{\frac{\beta}{2 \pi}} &  \frac{1}{2 \sqrt{\pi}} & \frac{1}{4 \sqrt{\pi \beta}} & \frac{1}{4 \sqrt{6 \pi} \beta} &  \frac{1}{16 \sqrt{3 \pi} \beta^{3/2}} & 
\frac{1}{16 \sqrt{30 \pi} \beta^2} & \frac{1}{96 \sqrt{10 \pi} \beta^{5/2}} \\ 
1 &  - \frac{1}{2 \sqrt{2 \pi \beta}} &  - \frac{1}{4 \sqrt{2 \pi} \beta} & 
- \frac{1}{8 \sqrt{3 \pi} \beta^{3/2}} & 
\frac{1}{16 \sqrt{6 \pi} \beta^2} &  \frac{1}{32 \sqrt{15 \pi} \beta^{5/2}}  &    &    \\ 
2 & \frac{1}{8 \sqrt{2 \pi} \beta^{3/2}} & 
\frac{1}{16 \sqrt{3 \pi} \beta^2} &  \frac{1}{32 \sqrt{6 \pi} \beta^{5/2}}  &    &    &    &    \\
3 &  -\frac{1}{48 \sqrt{2 \pi} \beta^{5/2}} &    &    &    &   
\end{array}
\end{align*}
\end{widetext}
This leads us to make an ansatz for the coefficients $C^{(m)}_{n, l}$
\begin{eqnarray}
C_{n,l}^{(m)} \xrightarrow{\beta \rightarrow \infty} \frac{(-1)^n}{\sqrt{n! (n+l)! \beta^{2n + l} 2^{2n + l}}} \sqrt{\frac{\beta}{2 \pi}}
\end{eqnarray} 
\\
which indeed satisfies the recursion relation (\ref{recursionInfty}).
Therefore the coefficients corresponding to $l = 0$ are given by
\begin{eqnarray}
C^{(2m)}_{m,0} \xrightarrow{\beta \rightarrow \infty} \sqrt{\frac{\beta}{2 \pi}} \bigg(- \frac{1}{2 \beta} \bigg)^m \frac{1}{m!}. \label{Cm0}
\end{eqnarray}
\begin{widetext}
Using Eq.~\eqref{Cm0} in the series solution of $\tilde{P}(\rho)$ we obtain
\begin{eqnarray}
\tilde{P}(\rho)
     &=& \frac{D_{\theta}}{D_{t}} e^{-\mu k \rho^2/2D_t} \sqrt{2 \pi \beta} \sum_m \lambda^{2m} \sqrt{\frac{\beta}{2 \pi}} \bigg(- \frac{1}{2 \beta} \bigg)^m \frac{1}{m!} L_m\bigg( \frac{\mu k \rho^2}{2 D_t} \bigg)^m \nonumber \\
     &=& \frac{\mu}{D_{t}} e^{-\mu k \rho^2/2D_t} \sum_m \bigg(-\frac{\lambda^2}{2\beta} \bigg)^{m} \frac{1}{(m!)}\sum_{k=0}^{m} \binom{m}{k} \frac{(-1)^k}{k!} \bigg( \frac{\mu k \rho^2}{2 D_t} \bigg)^k.
\end{eqnarray}
\end{widetext}
Evaluating the double sum after the following rearrangement $\sum_{m=0}^{\infty}\sum_{i=0}^m a_m b_i^m x^i = \sum_{m=0}^{\infty}\big(\sum_{i=m}^{\infty} a_i b_m^i) x^m$ 
and using $I_\alpha(x) = \sum_{m=0}^\infty \frac{1}{m!\, \Gamma(m+\alpha+1)}\big(\frac{x}{2}\big)^{2m+\alpha}$, the probability distribution in the limit $D_\theta \rightarrow 0$ becomes,
\begin{eqnarray}
\tilde{P}(\rho) = \frac{\mu k}{D_t} exp\bigg( - \frac{u_0^2}{2D_t k \mu} \bigg) I_0 \bigg( \frac{u_0 \rho}{D_t} \bigg)  e^{-\mu k \rho^2/2 D_t}~~~~
\end{eqnarray}
\\
which is Eq.~\eqref{dth-0-limit}. This is in accordance with Eq.~(8) of \cite{pototsky}.


\end{document}